\documentclass{mnras}
\usepackage{graphicx}

\def\pg{{PG1211+143}}
\def\mkn{{Mkn509}}

\def\xmm{{\it XMM-Newton}}

\def\suzaku{{\it Suzaku}}

\def\et{{et al.\ }}

\newcommand{\ls}{\mathrel{\hbox{\rlap{\hbox{\lower4pt\hbox{$\sim$}}}\hbox{$<$}}}}
\newcommand{\gs}{\mathrel{\hbox{\rlap{\hbox{\lower4pt\hbox{$\sim$}}}\hbox{$>$}}}}

\def\Msun{\hbox{$\rm ~M_{\odot}$}}

\def\rchi{{$\chi^{2}_{\nu}$}}

\def\H0{{\rm ~km~s^{-1}~Mpc^{-1}}}

\def\et{{et al.}}

\title [Ultra fast inflow]
       {Are ultrafast {\it inflows} in AGN truly rare - or just much harder to see?}
\author{Ken Pounds,
       Physics and Astronomy, University of Leicester, Leicester, LE1 7RH, UK\\}

\date{Accepted ; Submitted }
\pagerange{\pageref{firstpage}--\pageref{lastpage}}
\pubyear{2005}
\begin{document}

\label{firstpage}
\pagerange{\pageref{firstpage}--\pageref{lastpage}}

\maketitle

\begin{abstract}
Short-term variability and multiple velocity components in the powerful highly ionized wind of the archetypal UFO \pg\ are indicative of inner
disc instabilities or short-lived accretion events. The detection of a high velocity ($\sim$ 0.3c) {\it inflow} of highly ionized matter, located at 20 R$_{g}$,
offered the first direct observational support for the latter scenario, where matter approaching at a high inclination to the black hole spin plane may result in warping and
tearing of the inner accretion disc, with subsequent inter-ring collisions producing shocks, loss of rotational support and rapid mass infall. 
Simultaneous soft x-ray spectra reveal a lower velocity ($\sim$ 0.1c) inflow of less ionized matter, identified as 'upstream' at 200 R$_{g}$, with
a line of sight through matter converging on the supermassive black hole. We discuss here why ultrafast ionized winds are relatively common in luminous Seyfert galaxies, while detection of
the 0.3c inflow in \pg\ remains a rare example.
\end{abstract}

\begin{keywords}
galaxies: active -- galaxies: Seyfert: quasars: general -- galaxies:
individual: PG1211+143 -- X-ray: galaxies
\end{keywords}

\section{Introduction}
X-ray spectra from an \xmm\ observation of the luminous Seyfert galaxy \pg\ in 2001 first identified strongly blue-shifted
absorption lines of highly ionized gas, corresponding to a sub-relativistic outflow with a velocity of
0.15$\pm$0.01$c$ (Pounds \et\ 2003; Pounds \& Page 2006). A second example quickly followed with the detection of a powerful high velocity wind in PDS 456 (Reeves \et\ 2003), with archival data
from \xmm\ and \suzaku\ subsequently showing that such ultra-fast,
highly-ionized outflows (UFOs) are relatively common in nearby, luminous AGN (Tombesi \et\ 2010, 2011; Gofford \et\ 2013). While those archival
searches typically reported a single velocity, in the few cases where an AGN was observed repeatedly, the wind velocity was often different.
\mkn\ is the best example from the \xmm\ data archive, with wind velocities of $\sim$0.173c, $\sim$0.139c and $\sim$0.196c, separated by 5 years
and 6 months respectively (Cappi \et\ 2009).

\begin{figure*}                                 
\centering                                                              
\includegraphics[width=7cm,angle=-90]{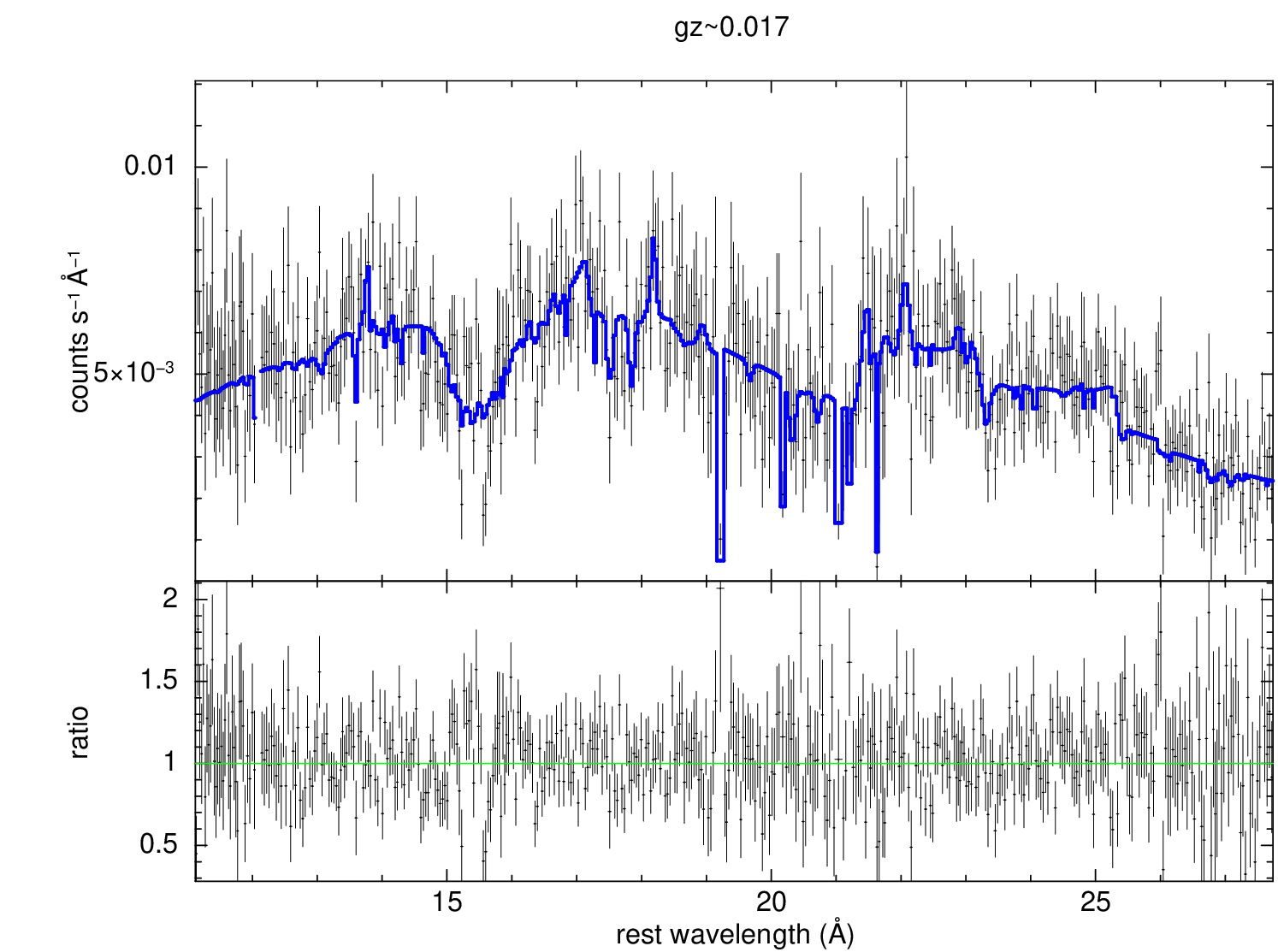}
\includegraphics[width=7cm,angle=-90]{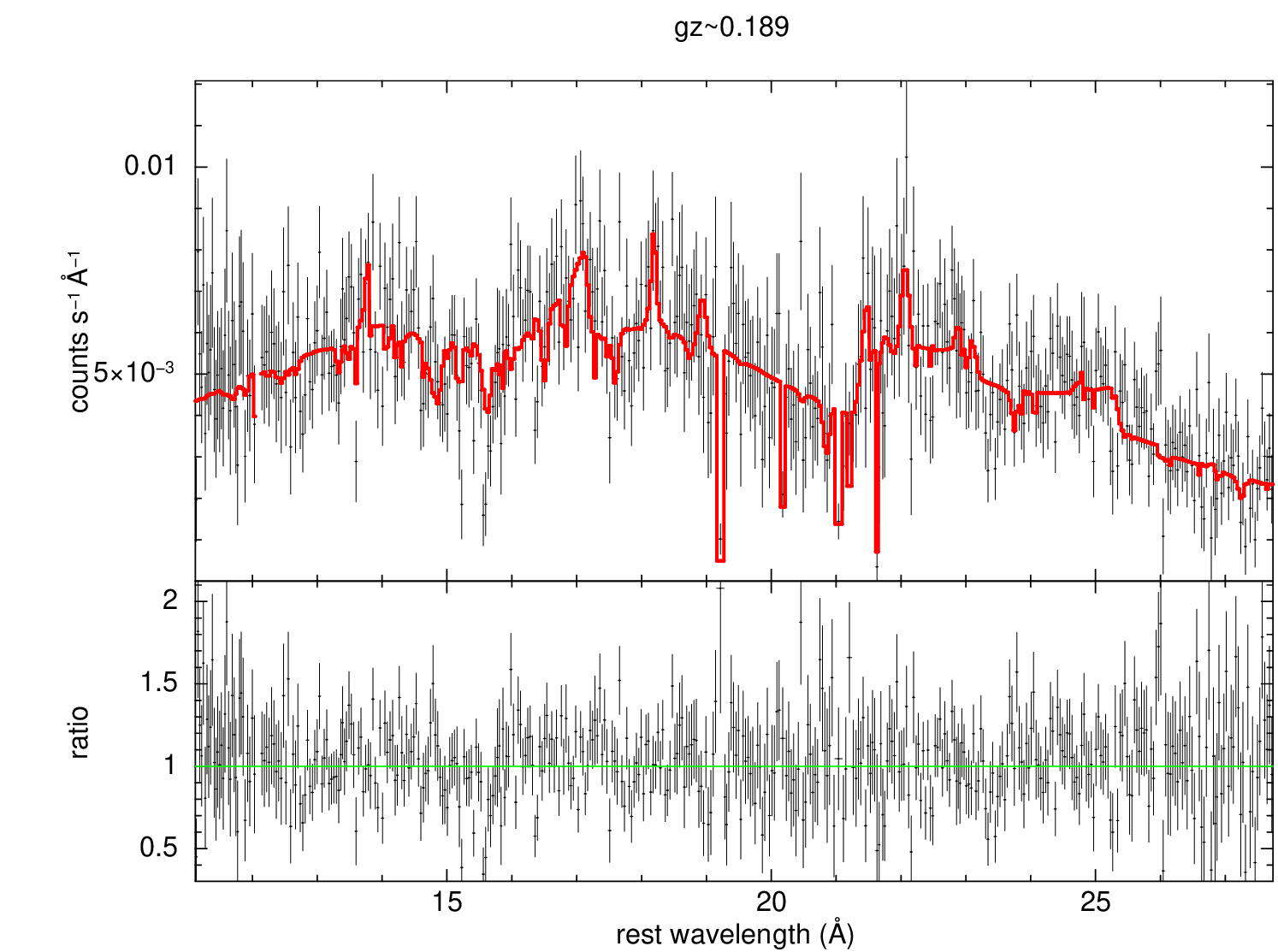}
\includegraphics[width=7cm,angle=-90]{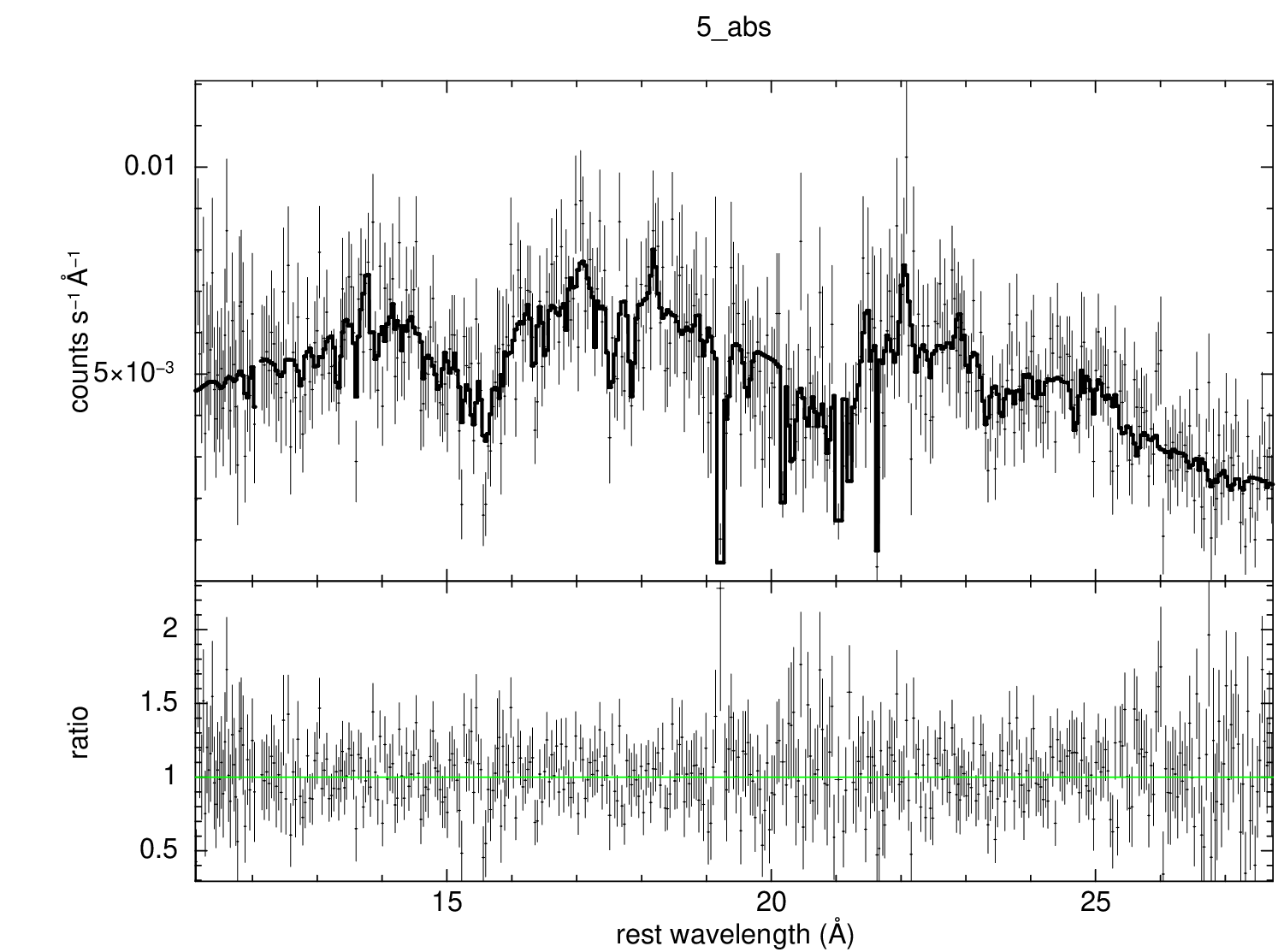}
\caption{rev2659 soft x-ray spectral fit with continuum and photoionized emission features from the mean 2014 spectrum with the addition of: (top panel) blue-shifted absorption corresponding to a prevailing outflow
velocity v$\sim$0.061c;
(mid-panel) red-shifted absorption from inflowing matter at v$\sim$0.096c; (lower panel) combined blue- and red-shifted absorption components yielding an excellent overall spectral fit.}  
\end{figure*}

An extended 5-weeks \xmm\ observation of \pg\ in 2014,covering 7 spacecraft orbits, showed a more complex velocity structure in the highly ionized
wind, with three primary (high column density)  outflow velocities of $v \sim 0.066c$, $v \sim 0.13c$ and $v \sim 0.19c$
(Pounds  \et\ 2016a), none being consistent with the wind velocity observed in 2001.  In a review of continuum-driven or
'Eddington' winds, King and Pounds (2015) furthermore showed that the observability of an individual, short-lived wind ejection may be of order months or
less, as an expanding shell presents a lower absorbing column to the nuclear x-ray source.

Considering the simultaneous observation of multiple expanding shells of absorbing gas in the context of super-Eddington accretion, Pounds,
Lobban  and Nixon (2017) noted that short-lived wind components might be a natural consequence of the way in which matter accretes in an AGN,
typically falling from far outside the sphere of gravitational influence of the SMBH and with essentially random orientation. The resulting 
accretion stream will in general orbit in a plane misaligned to the spin of the central black hole (King \& Pringle 2006, 2007), with the inner
disc subject to Lense-Thirring precession around the spin vector, and orbits at smaller radii precessing sufficiently fast to cause
the tearing-away of independent rings of gas.

Computer simulations by Nixon \et\ (2012) furthermore showed that - as each torn-off ring precesses on its own timescale - two neighbouring rings will eventually
collide, with the shocked material losing rotational support and falling inwards to a new radius defined by its residual angular momentum. A substantial increase in the
local accretion rate might then allow a previously sub-Eddington flow to become briefly super-Eddington, with excess matter being ejected as a wind at or above
the local escape velocity, as predicted in the classic paper on accretion disc theory by Shakura and Sunyaev (1973). For typical AGN disc parameters, Nixon \et\ (2012)
show the expected `tearing radius' is of order a few hundred gravitational radii from the black hole - with predicted (escape) wind speeds in the range observed (King and Pounds 2015;
fig 4a).

The detection of red-shifted X-ray absorption spectra during part of the same 2014 \xmm\ observation of \pg, corresponding to 
a substantial {\it inflow} of matter approaching the black hole at velocities up to
$\sim$ 0.3c (Pounds \et\ 2018; hereafter Po18), offered the first observational support for the above accretion scenario. Simultaneous hard and soft X-ray spectra provided
independent confirmation of the high inflow velocity, with the soft X-ray absorption indicating
less ionized (higher density) matter embedded in the primary highly ionized flow. No further confirmed examples of an ultrafast AGN inflow have been reported, at least to this author's knowledge,
which is surprising in light of the above discussion. As noted in Po18, the probability of detecting such an extreme velocity inflow, close to the black hole, will be reduced both by a lower
chance of a narrow converging flow being in the line of sight to the continuum source, and the likely short duration of the infall. Such considerations suggest a lower velocity, less highly ionized
inflow, such as that also seen in \pg,  could offer a better opportunity for detection

In the present paper we report a new analysis of the soft x-ray data from the Reflection Grating Spectrometer (RGS; den
Herder \et 2001) in the 2014 \xmm\ observation of \pg, separately identifying blue- and red-shifted absorption components. In addition
to the relatively weak high redshift absorber (at $z=0.43$), we confirm a
stronger soft x-ray absorber at a redshift $z=0.189$, corresponding to an inflow velocity
$\sim$0.1c, and (free-fall) location at 200 R$_{g}$. We identify this absorption with matter lying 'upstream' of the ultrafast, highly ionized inflow reported in Po18, in the context
of a line of sight along an accelerating flow converging on the black hole. 

We assume a cosmological redshift for \pg\ of $z=0.0809$ (Marziani \et\ 1996), with a black hole mass of $4\times 10^{7}$\Msun\ (Kaspi \et\
2000) indicating the historical bolometric luminosity is close to Eddington. All velocities are corrected for the relativistic Doppler effect.
The spectral analysis is based on the software package described in {\tt XSPEC} (Arnaud \et\ 1996.

\section{Soft x-ray absorption coincident with the hard x-ray inflow in \pg.}

In order to quantify the strong red-shifted absorption unique to the second orbit (rev2659) of the 2014 campaign, we start with the mean 2014 soft x-ray spectral profile of \pg\ published in
Pounds \et 2016b (Po16b), based on the sum of RGS1 and RGS2 data from all 7 \xmm\ orbits (for a combined exposure of 1.27 Ms). Fitting the data over the 11-28 \AA\ waveband, where the data are of
highest statistical
quality, the continuum is modelled by power law and black body components, both attenuated by the Galactic column.
Strong positive and negative residuals 
were matched in Po16b with photoionized emission and absorption attributed to a circumstellar outflow. Details of the multiplicative absorption and additive emission
grids used are also given in Po16b.

Here, we substitute the composite 7-orbit RGS data with that from rev2659, and repeat the spectral fit, retaining continuum and photoionised emission,
with only the normalisation of each component left free. The blue-shifted absorption reported in Po16b is initially excluded, though we expect it to be present at some level throughout all 7
orbits.
The fit is now quite poor, with \rchi\ of 554/436 over the 11-28 \AA\ waveband.

In a second spectral fit (Figure 1; top panel), blue-shifted absorption is added, finding the primary (high column) outflow velocities of $\sim0.06c$ and $\sim0.13$ reported in Po16b, with 
column density and ionization parameters listed in the Table 1. This addition recovered a much improved fit, with \rchi\ of 485/432.   

The data from rev2659 are next examined in the same way for red-shifted absorption. All 3 components mentioned in Pounds \et\ 2018 are found, with column density, ionization parameter and
redshifts
listed in Table 1, together with their individual significance. The fit statistic has \rchi\ of 525/427, weaker than for the blue-shifted absorption alone, but
still highly significant.

Finally, re-fitting the 7-orbit baseline emission model with both red- and blue-shifted absorption yields an excellent spectral fit (\rchi\ of 455/421), with the
combined flows provide a better match, in particular to the complex of inner-shell Fe absorption lines (or UTA; eg Behar \et\ 2001).   
For visual clarity in Figure 1 we confine the blue- and red-shifted spectral plots to the dominant flow component in each case.

\begin{table*}
\centering
\caption{Parameters of 3 red- and 2 blue-shifted soft x-ray absorption components from a spectral fit to the RGS data from rev2659. Each absorber is further defined by its ionisation parameter 
$\xi$ (erg cm s$^{-1}$) and column density N$_{H}$ (cm$^{-2}$). }
\begin{tabular}{@{}lcccc@{}}
\hline
log$\xi$ & N$_{H}$ (10$^{20}$) & redshift & v$_{in}$  & $\Delta\chi^{2}$ \\
\hline
2.58$\pm$0.15 & 11$\pm$4  & 0.189$\pm$0.001 & 0.095$\pm$0.001 & 17/3 \\
3.42$\pm$0.18 & 40$\pm$43 & 0.172$\pm$0.001 & 0.080$\pm$0.001 &  12/3 \\
0.76$\pm$0.23 & 8$\pm$4 & 0.433$\pm$0.002 & 0.28$\pm$0.01 &  9/3 \\
0.85$\pm$0.18 & 4.2$\pm$2.1 & -0.052$\pm$0.001 & -0.131$\pm$0.001 &  13/3 \\
1.32$\pm$0.08 & 8$\pm$4 & -0.0168$\pm$0.0005 & -0.0611$\pm$0.0005 &  65/3 \\
\hline
\end{tabular}
\end{table*}

In the context of the present paper we speculate that previous analyses of the soft x-ray spectra may well have missed a short-lived inflow in the presence of a stronger, more persistent outflow.
We acknowledge that was the case in our analysis of the variable soft x-ray wind in \pg\ (Reeves \et\ 2018), carried out before the highly ionised inflow detection at 0.3c was made. 

\section{Discussion}
The discovery of ultra-fast highly ionized outflows (UFOs) in early x-ray spectra of the narrow line
Seyfert galaxy \pg\ (Pounds \et\ 2003) and
the luminous QSO PDS 456 (Reeves \et\ 2003) opened up a new field of study of AGN, well suited to the uniquely high throughput of x-ray
spectrometers on ESA's \xmm, launched in late 1999. King and Pounds (2003, 2015) noted that such winds are a natural result of a high accretion ratio,
with excess matter being driven off by radiation pressure when the accretion rate exceeds the local Eddington limit (Shakura and Sunyaev
1973).

While this picture provided a satisfactory explanation of most UFOs, where a single detection yielded a unique wind velocity, the extended
study of \pg\ in 2014 found a more complex outflow profile, with velocities of $\sim$0.06c, $\sim$0.13c and $\sim$0.18c detected in the
stacked data set. Such complexity was clearly inconsistent with a wind profile launched from a flat axi-symmetric disc (SS73), suggesting some intrinsic disc
instability or rapidly variable accretion rate  being introduced to the inner disc.

The detection of red-shifted X-ray absorption spectra during part of the same 2014 \xmm\ observation of \pg, where the extreme redshift ($\sim$0.48) in highly ionized matter
indicated absorption
in a substantial {\it inflow} approaching the black hole at a velocity of
$\sim$ 0.3c (Po18), offered the first observational support for such rapidly rapidly variable accretion.
While the inflow observation was of high significance in only one of seven individual observations, inflows of lower column density and smaller redshift were detected in 5 of the
other 6 EPIC pn observations, with redshifts ranging from 0.20 to 0.36 (and line-of-sight inflow velocities from 0.1 -- 0.23c).    

As noted in the Introduction, periods of highly variable accretion to the inner disc are a likely 
consequence of the way in which AGN accrete, where gas initially falls towards the SMBH with essentially random orientation. 
Lense-Thirring precession would cause misaligned orbits to precess around the black hole spin vector, with the innermost ring(s) warping and potentially breaking
off.  Collision between 
neighbouring rings, rotating at different rates, would then shock, with loss of rotational support leading to matter falling freely to a new radius defined by
its residual  angular
momentum, where it may form a new disc (Nixon \et\ 2012).

Our new soft x-ray analysis, reported here, shows both blue- and red-shifted photionized absorption strongly affecting the soft x-ray spectrum of \pg\ during the 2014 \xmm\ observations.
The higher significance of the outflow components is in part due to the 7-fold greater exposure for a wind that persists throughout the \xmm\ observation,
with the outflow at $\sim0.061c$ particularly strong (see also Reeves \et 2018). In contrast, red-shifted absorption is limited to rev2659 and much diluted in the overall 2014 data.

We identify the soft x-ray inflow component at $\sim0.433c$ with higher density matter surviving close to the primary, highly ionised inflow reported in Po18. While providing independent support
of the primary $\sim$0.3c inflow, the accretion mass rate remains dominated by the highly ionised flow.

The strongest soft x-ray inflow component, at $\sim$0.095c, is the most interesting and here we identify that slower moving and less highly ionized absorber with matter
'upstream' from the 0.3c
inflow, with the correlation of ionization parameter and velocity being also indicative of a converging and accelerating inflow approaching the SMBH. In this impoortant respect we  differ from
Reeves \et\ (2018), where the absorption in rev2659 was discussed as a random cloud or filament crossing the line of sight.  
Assuming, as before, a free-fall velocity determined solely by
the black hole's gravity, the strong soft x-ray inflow is located at 200 R$_{g}$.

Splitting the $\sim$100ks observation into 5 equal time intervals finds both high and low ionisation inflows present at similar strength throughout rev2659, showing the flow mass rate is
maintained throughout the rev2659 observation. However, the high ionization, high column inflow is not detected in either preceding (rev2652) or following (rev2661) spacecraft orbits. 

Po18 estimate an accretion rate during rev2659 for the highly ionized inflow of $\sim 10^{23}$ gm s$^{-1}$, yielding a luminosity of $\sim 10^{43}$ erg s$^{-1}$
for a mass/energy conversion efficiency $\eta\sim$0.1, and lasting for $\sim 10^{5}$ s. 
In comparison, the mean bolometric luminosity of \pg\ is $\sim 10^{45}$ erg s$^{-1}$, with $\sim 10^{44}$ erg s$^{-1}$ arising from the inner disc region relevant here,
suggesting a small number of
infall events - similar to that observed - would make a significant contribution to the inner disc accretion rate, while remaing out of sight.

While the lower column density of a soft x-ray counterpart of the 0.3c inflow is of less significance to the overall accretion rate, detection of co-moving,
higher density matter in the higher resolution RGS spectra provides valuable confirmation of that extreme velocity. In addition, the strongest red-shifted absorber provides
an upstream measure of an accelerating stream of matter as it converges and becomes more highly ionized on close approach to the SMBH in \pg.
Most relevant in the context of the present paper, the slower and broader upstream inflow is likely to be easier to detect. A search for such soft x-ray inflows in new or archival \xmm\
observations is highly recommended.

\section{Data availability}

The data underlying this article are available in the XMM archive at http://nxsa.esac.esa.int/nxsa-web.

\section*{ Acknowledgements }
\xmm\ is a space science mission developed and operated by the European Space Agency. We acknowledge in particular the excellent
work of ESA staff in Madrid in successfully planning and conducting the \xmm\ observations. I am particularly grateful for the collaboration of Andrew Lobban who provided the initial data reduction
on which this note is based,
as well as being an outstanding colleague in researching the x-ray properties of \pg\ over many years.

\bsp	
\label{lastpage}

\end{document}